\documentclass[nofootinbib,showpacs,floatfix,onecolumn,prd]{revtex4-1}
\usepackage{graphicx,psfrag,amsmath,amssymb,amsfonts,latexsym,color,epsf,dcolumn,graphpap}
\usepackage{subfig}
\usepackage{float}
\usepackage{lipsum}
\usepackage{enumerate}
\usepackage{cases}

\usepackage{hyperref}
\hypersetup{colorlinks=true,linktoc=page,linkcolor=blue,citecolor=red,urlcolor=cyan}

\usepackage{tikz-cd} 
\usepackage{tikz}
\usetikzlibrary{shapes.geometric, arrows}

\begin{document}
\title{Nonlocal effective action and particle creation in $D$ dimensions}
\author{Andr\'es Boasso$^{1,2}$}
\author{Sebastián Franchino Viñas$^{3,4,5}$}
\author{Francisco D. Mazzitelli$^{1,2}$}
\affiliation{$^1$Centro At\'omico Bariloche,
Comisi\'on Nacional de Energ\'\i a At\'omica, 
R8402AGP Bariloche, Argentina}
\affiliation{$^2$Instituto Balseiro,
Universidad Nacional de Cuyo,
Centro Atómico Bariloche, 
R8402AGP Bariloche, Argentina}
\affiliation{$^3$ DIME, Universit\`a di Genova, Via all'Opera Pia 15, 16145 Genova, Italy}
\affiliation{$^4$ INFN Sezione di Genova, Via Dodecaneso 33, 16146 Genova, Italy}
\affiliation{$^{5}$
Laboratoire d’Annecy-le-Vieux de Physique Théorique, CNRS – USMB, BP 110 Annecy-le-Vieux,
74941 Annecy, France}

\begin{abstract}
We compute the particle creation rate in the context of quantum fields in curved spacetimes, by evaluating the imaginary part of the effective action up to second order in the curvatures. For arbitrary metrics in dimensions $D\geq4$, we express the vacuum persistence amplitude in terms of the Ricci scalar and the Weyl tensor, showing that, up to their second power, no particle creation occurs for conformal fields in conformally flat spacetimes. We pinpoint an analogy with the electromagnetic pair creation, by writing the squared Weyl tensor invariant in terms of its electric and magnetic parts. In addition, we present an alternative expression for the imaginary part of the effective action employing the Cotton tensor. This is particularly useful in $D=3$, where the Weyl tensor trivially vanishes and the Cotton tensor is related to conformal flatness. Finally, we highlight the importance of the threshold for particle creation, a point that has been overlooked in some recent studies. 
\end{abstract}

\date{today}


\maketitle
\section{Introduction}
Particle creation is of utmost importance in several branches of physics, including  quantum electrodynamics in the presence of strong background fields \cite{Schwinger} and quantum field theory in curved spacetimes \cite{Birrel}. 
Generally speaking, there are two main approaches to analyze it: the first one, originally described in Parker's seminal work \cite{Parker68}, consists in the calculation of the  Bogoliubov transformation that connects the $\vert 0_{\text{in}}\rangle$ and $\vert 0_{\text{out}}\rangle$ vacua, while the second  one is based on the evaluation of the imaginary part of the effective action, which is related to the vacuum persistence probability $\vert\langle 0_{\text{in}}\vert 0_{\text{out}}\rangle\vert^2$.

The main ingredient for pair creation from the first point of view is the mixing, as time flows, of the positive and negative frequency modes in the presence of a time-dependent background \cite{Birrel}. Notice that the in and out vacua (or more generally the vacua at different times) are well defined only during static or adiabatic periods~\cite{Castag85}; otherwise, even the states that minimize the Hamiltonian at a given time are in general not admissible states, because they give rise to two-point functions that do not reproduce Hadamard singularities, and therefore prevent a covariant renormalization of the theory \cite{Castag87}. In the particular case of static configurations, the existence of a Killing vector field unambiguously defines  the notion of positive and negative frequency modes at any time, and therefore no particle creation occurs. However, this is of course not the case for static backgrounds that produce unstable modes, like a static electric field  or a position-dependent mass that becomes negative in some spatial regions.

On the other hand, the heat kernel technique \cite{Vassilevich} has proven to be a useful tool for the computation of effective actions in the presence of background fields.  In its traditional application in physics, the heat kernel is introduced using an expansion in powers of the proper time $\tau$. The term proportional to $\tau^n$ involves the  Schwinger--DeWitt coefficient $a_n(x,x')$, which is a function of the background fields and derivatives. In this way, the result is an expansion of the effective action in inverse powers of the mass of the field. More precisely, such a result is valid as long as $\nabla\nabla \mathcal{R} \ll m^2 \mathcal{R}$, where 
$\mathcal{R}$ denotes generic components of the Riemann tensor. In this original approach, which is not valid for massless fields,  the effective action is real and, consequently, particle creation can not be observed.

Instead, as noted by Vilkovisky many years ago \cite{Gospel}, for light or massless fields one can use a different expansion, valid for fastly varying fields, i.e.~$\nabla\nabla\mathcal{R}\gg \mathcal{R}^2$. This technical development involves a resummation of an infinite number of terms in the propertime expansion of Seeley--DeWitt; ordinarily, this gives access to nonlocal contributions to the effective action and the corresponding imaginary parts~\cite{BarviVilko, Avramidi}. 

In this paper, we compute the vacuum persistence probability in an arbitrary number of dimensions $D$, using the above-described nonlocal effective action (expanded up to second order in the curvature). We generalize previous results obtained for $D=4$~\cite{Frieman,Maroto,Campos94} and explore general properties, with particular focus on massless fields. Additionally, we demonstrate that the effective action, typically expressed in terms of the  Ricci scalar and the Ricci (or sometimes the Weyl) tensor, can be reformulated trading the latter for the Cotton tensor. This reformulation will enable us to shed light on aspects of conformal symmetry in $D=3$.

The paper is organized as follows. In Sec.~\ref{sec:NL}, we describe the model under consideration: a scalar field with a spacetime-dependent mass and an arbitrary nonminimal coupling to the curvature scalar. In parallel, we also briefly review the results for the effective action up to second order in the curvature using the covariant perturbation theory of heat kernels. In Sec.~\ref{sec:imaginary}, we compute the imaginary part of the effective action in an arbitrary number of dimensions $D \geq 3$. For the special case of massless fields in $D \geq 4$, we show that the imaginary part can be expressed in terms of the Weyl tensor, emphasizing the absence of particle creation for conformal fields at quadratic order in the curvature. Additionally, for $D=4$ we express the vacuum persistence probability in terms of the electric and magnetic components of the Weyl tensor, drawing a direct analogy to the particle creation rate in scalar and spinor QED;  we also analyze in $D=4$ the pair creation process for an oscillating Newtonian star. As the Weyl tensor vanishes for $D=3$, Sec.~\ref{sec:cotton} is devoted to the reformulation of the effective action in terms of the Cotton tensor (we remind that the Cotton tensor plays the role of the Weyl tensor in $D=3$, in the sense that a 3-dimensional metric is conformally flat iff the Cotton tensor vanishes).
In Sec.~\ref{sec:threshold} we stress the relevance of the particle creation threshold in the imaginary part of the effective action. Although this is well known,  we include this discussion in order to clarify some recent claims in the literature regarding the eventual existence of particle creation for static gravitational backgrounds \cite{Wondrak,Chernodub}. 
Subsequently, in Sec.~\ref{sec:discussion},  we summarize and discuss the results obtained in this work.
In App.~\ref{app:geometric}, we include important identities regarding the geometric quantities that we use in this manuscript.

\subsection*{Conventions}
We use Planck units, the mostly minus convention for the metric in the Lorentzian setup, i.e. with signature $(-,+,+\cdots)$, and the definition $R_{\mu\nu}:=R^{\rho}{}_{\mu\rho\nu}$ for the Ricci tensor. As usual, Greek indices ($\mu,\nu, \cdots$) run over spacetime indices, while Latin indices ($a,b,\cdots,i,j,\cdots$) correspond just to the spatial sector.

\section{The nonlocal effective action}\label{sec:NL}

We consider a massive, real scalar field $\phi$, defined on an curved spacetime background of dimension $D$, with a spacetime-dependent squared mass and a nonminimal coupling $\xi$ to the scalar Ricci curvature $R$. 
For convenience, the mass is written explicitly in terms of a constant ($m^2$) contribution and a fluctuating ($\sigma^2$) one.
In Euclidean space the action reads 
\begin{equation}
	S_{\text{E}} := \frac{1}{2} \int\mathrm{d}^D\mathbf{x}\, \sqrt{g}\; \phi\left(-\square_{\text{E}} + m^2 +\sigma^2 + \xi R\right)\phi \,  ,
	\label{ec:S_euclidea}
\end{equation}
where $\square_{\text{E}}$ is the Euclidean  d'Alembertian.
In its turn, the Euclidean effective action obtained after the integration of the quantum scalar field $\phi$ is 
\begin{equation}
	\Gamma_{\text{E}} = \frac{1}{2}\operatorname{Log} \operatorname{Det}\left(-\square_{\text{E}} + m^2 +\sigma^2 + \xi R\right).
	\label{ec:S_eff_det}
\end{equation}
Assuming regularity, it can be computed using a covariant expansion in powers of the curvatures.
For simplicity we will assume that the spacetime-dependent part of the mass satisfies 
$\sigma^2=O(R)$.
The nonlocal part of the effective action, up to second order in the curvatures, is thus given by~\cite{BarviVilko,Avramidi}
\begin{multline}
	\Gamma^{(2)}_{\text{E}} := \frac{1}{2(4\pi)^{D/2}} \int \mathrm{d}^D\mathbf{x}\,\sqrt{g} \left[\sigma^2\beta^{(1)}(\square_E)\sigma^2 - 2\sigma^2\beta^{(2)}(\square_E)R + \xi \left(\sigma^2\beta^{(1)}(\square_E)R + R\beta^{(1)}(\square_E)\sigma^2\right)\right. \\
	\left.+R\left(\xi^2\beta^{(1)}(\square_E)-2\xi\beta^{(2)}(\square_E)+\beta^{(4)}(\square_E)\right)R + R_{\mu\nu}\beta^{(3)}(\square_E)R^{\mu\nu}\right],	\label{ec:Avramidi_euclidea}
\end{multline}
where $R_{\mu\nu}$ is the Ricci tensor. For massless fields, the functions $\beta^{(i)}$  involve $\log[-\square_E]$ and $[-\square_E]^{1/2}$, respectively for even and odd dimensions, what can be seen from their explicit expressions. Indeed, they read, for $D>2$,
\begin{numcases}
	{\beta^{(i)}(\square_E) 
    :=
    \frac{(-1)^{\frac{D}{2}}}{2\Gamma\left(\frac{D}{2}-1\right)} \int_{0}^{1}\mathrm{d}z\, f^{(i)}(z) \left(m^2 - \gamma \square_E \right)^{\frac{D}{2}-2} 
    }
    \log\left(\frac{m^2-\gamma \square_E}{\mu^2}\right)\, , \quad \text{D even},
	\label{ec:Avramidi_beta_par}
    \\
    (-1)^{-\frac{1}{2}}\pi\, , \quad \text{D odd},
	\label{ec:Avramidi_beta_impar}
\end{numcases}
where $\mu$ is a scale of mass dimensions introduced by dimensional regularization.
We have additionally defined 
\begin{align}
    \gamma:=\frac{1-z^2}{4},
\end{align}
and the functions $f^{(i)}$ are given by
\begin{align}
	\begin{split}
	f^{(1)}(z) :&= 1,  \\
	f^{(2)}(z) :&= \gamma, \\
	f^{(3)}(z) :&= \frac{z^4}{6},\\
	f^{(4)}(z) :&= \frac{1}{48} \left(3-6z^2-z^4\right).
    \end{split}
 \label{ec:efes}
\end{align}
In what follows we will use these results to compute the imaginary part of the Lorentzian effective action and will associate it to the total pair production probability $P$.

\section{Imaginary part of the Lorentzian effective action}\label{sec:imaginary}
To compute the vacuum persistence probability, we will evaluate the in-out effective action in Lorentzian signature. This calculation can be approached in various ways. A direct method is to consider weak gravitational fields
\begin{align}\label{eq:expansion}
    g_{\mu\nu}=\eta_{\mu\nu}+ h_{\mu\nu},
    \quad h_{\mu\nu} \ll1,
\end{align}
compute the effective action up to second order in $h_{\mu\nu}$
and covariantize the final result, recasting all the contributions involving $h$ in terms of geometric tensors. This calculation has been done in detail in Ref.~\cite{Campos94}, both for the in-in and the in-out effective actions. The result for the in-out effective action coincides with the Euclidean one after a Wick rotation of the (flat) Euclidean propagators. We will use here this shortcut, making the following substitutions in the Euclidean effective action:
$\square_{\text{E}}\mapsto\square$ and  $m^2\mapsto m^2-i\epsilon$. 

Afterwards, recall that the relation between the vacuum persistence probability and the imaginary part of the Lorentzian effective action is
\begin{equation}
	\left|\langle 0_{\text{out}}|0_{\text{in}}\rangle\right|^2 = e^{-2\text{Im}\left(\Gamma\right)} = 1 - 2\text{Im}\left(\Gamma\right) + \mathcal{O}\left(\mathcal{R}^3\right) ,
	\label{ec:probabilidad}
\end{equation}
so that, for weak fields, the pair production probability is
$P\simeq 2\text{Im}\left(\Gamma\right).$

To compute the imaginary part of the effective action, we note that, in odd dimensions, the form factors $\beta^{(i)}$ in Eq.~\eqref{ec:Avramidi_beta_impar} are made of polynomials of $\square$ times a function with a branch cut in the real negative axis:
\begin{equation}
	\left(m^2 - i \epsilon - \gamma \square \right)^{\frac{D}{2}-2} = \left(m^2-i\epsilon - \gamma \square \right)^{\frac{D-3}{2}}\left(m^2 - i \epsilon - \gamma \square \right)^{-1/2}.
	\label{ec:pot_semientera}
\end{equation}
Similarly, for even dimensions we observe the presence of a logarithmic branch cut in Eq.~\eqref{ec:Avramidi_beta_par}:
\begin{equation}
	\left(m^2-i\epsilon - \gamma \square \right)^{\frac{D}{2}-2}\log\left(\frac{m^2-i\epsilon-\gamma \square}{\mu^2}\right)\, .
	\label{ec:logaritmo}
\end{equation}

The imaginary part of the different form factors is easily computed in Fourier space, by expanding around a flat metric. Indeed, when acting on a real test function $\psi\left(\mathbf{x}\right)$, in odd dimensions and flat space we have\footnote{We will use an abuse of notation, denoting both a function and its Fourier trasnform with the same symbol.}
\begin{equation}
	\int\mathrm{d}^D\mathbf{x}\,\psi(\mathbf{x})\left(m^2 -i\epsilon - \gamma\,\square \right)^{-1/2}\psi(\mathbf{x}) = \int\frac{\mathrm{d}^D\mathbf{p}}{(2\pi)^D}\, \vert\psi\left(\mathbf{p}\right)\vert^2\, \left[m^2-i\epsilon + \gamma\mathbf{p}^2 \right]^{-1/2}.
	\label{ec:semientera_2}
\end{equation}
Therefore, in the limit $\epsilon\rightarrow 0^{+}$ we obtain
\begin{equation}
\text{Im}\left[\int\mathrm{d}^D\mathbf{x}\,\psi(\mathbf{x})\left(m^2-i\epsilon - \gamma\,\square \right)^{-1/2}\psi(\mathbf{x})\right] = \int\frac{\mathrm{d}^D\mathbf{p}}{(2\pi)^D}\, \vert\psi\left(\mathbf{p}\right)\vert^2 \, \left(-m^2 - \gamma\,\mathbf{p}^2\right)^{-1/2}\,\Theta\left(-m^2-\gamma\,\mathbf{p}^2\right) \, .\label{ec:imaginaria_semientera}
\end{equation}
A similar procedure for a flat spacetime of even dimensions gives 
\begin{equation}
\text{Im}\left[\int\mathrm{d}^D\mathbf{x}\,\psi(\mathbf{x})\log\left(m^2-i\epsilon - \gamma\,\square \right)\psi(\mathbf{x})\right] \\
	= -\pi \int\frac{\mathrm{d}^D\mathbf{p}}{(2\pi)^D}\, \vert\psi\left(\mathbf{p}\right)\vert^2\,\Theta\left(-m^2-\gamma\,\mathbf{p}^2\right) .
\label{ec:imaginaria_logaritmo}
\end{equation}

Using these results, we can now compute the imaginary part of the nonlocal effective action. In doing so, it is useful to note that
\begin{equation}
\int_0^1 \mathrm{d}z\, f(z)\Theta\left(-m^2-\gamma\,\mathbf{p}^2\right)\cdots\,  =
\int_{0}^{z_{\text{max}}}\mathrm{d}z\, f(z)\Theta\left(-4 m^2-\mathbf{p}^2\right)\cdots\, ,
\label{ec:intz}
\end{equation}
with $z_{\text{max}}:=\sqrt{1+4m^2/\mathbf{p}^2}$, which renders explicit the threshold for the creation of a particle-antiparticle pair.  Inserting  Eqs. \eqref{ec:semientera_2}-\eqref{ec:intz} into Eqs. \eqref{ec:Avramidi_euclidea}-\eqref{ec:efes}
we obtain, for both even and odd $D$, 
\begin{multline}
	P = \frac{\pi^{1-D/2}}{2\Gamma\left(\frac{D}{2}-1\right)} \int\frac{\mathrm{d}^D\mathbf{p}}{(4\pi)^D}\,\Theta\left(-\mathbf{p}^2 - 4m^2\right) \\
    \times
\left(\alpha^{(1)}\sigma^2\left(-\mathbf{p}\right)\sigma^2\left(\mathbf{p}\right)+\alpha^{(2)}R\left(-\mathbf{p}\right)\sigma^2\left(\mathbf{p}\right)+\alpha^{(3)}R\left(-\mathbf{p}\right)R\left(\mathbf{p}\right)+\alpha^{(4)}R^{\mu\nu}\left(-\mathbf{p}\right)R_{\mu\nu}\left(\mathbf{p}\right)\right) ,
	\label{ec:proba_final}
\end{multline}
where the $\alpha^{(i)}$ functions are defined as 
\begin{align}
	\alpha^{(1)} :&=  \int_{0}^{z_{\text{max}}}\mathrm{d}z\, \left(-m^2-\gamma\mathbf{p}^2\right)^{\frac{D}{2}-2} ,\notag\\
	\alpha^{(2)} :&= 2\int_{0}^{z_{\text{max}}}\mathrm{d}z\, \left(-m^2-\gamma\mathbf{p}^2\right)^{\frac{D}{2}-2} \left(- \gamma + \xi\right), \notag\\
	\alpha^{(3)} :&= \int_{0}^{z_{\text{max}}}\mathrm{d}z\, \left(-m^2-\gamma\mathbf{p}^2\right)^{\frac{D}{2}-2} \left(\xi^2-2\xi \gamma+\frac{1}{48}\left(3-6z^2-z^4\right)\right) ,\notag\\
	\alpha^{(4)} :&= \frac{1}{6} \int_{0}^{z_{\text{max}}}\mathrm{d}z\, \left(-m^2-\gamma \mathbf{p}^2\right)^{\frac{D}{2}-2} z^4 .
	\label{ec:alpha_mneq0}
\end{align}

\subsection{The massless case}

Computing the functions 
 $\alpha^{(i)}$ in the particular case 
 of a completely massless theory, i.e. setting $m=\sigma=0$, and 
 recalling that the conformal coupling in $D$ dimensions is $\xi_D := \frac{D-2}{4\left(D-1\right)}$, one can readily obtain an explicit expression for the probability of pair creation:
\begin{multline}
	P = \frac{\pi^{3/2} (4\pi)^{-D/2}}{2^{D+1}\Gamma\left(\frac{D+3}{2}\right)} \int\frac{\mathrm{d}^D\mathbf{p}}{(2\pi)^D}\,\Theta\left(-\mathbf{p}^2\right)\left(-\mathbf{p}^2\right)^{\frac{D}{2}-2}\Bigg[ R^{\mu\nu}\left(-\mathbf{p}\right)R_{\mu\nu}\left(\mathbf{p}\right) \Bigg. \\
	\left.+\left( 2\left(D^2-1\right)\left(\xi - \xi_D\right)^2 - \frac{D}{4\left(D-1\right)} \right)
	R\left(-\mathbf{p}\right)R\left(\mathbf{p}\right)\right].
	\label{ec:proba_final_m0}
\end{multline}

For physical reasons it is interesting to rewrite the particle creation probability in terms of the Weyl tensor. To do that, we use the fact that the usual Gauss--Bonnet identity, valid for $D=4$, remains valid in any number of dimensions when considered up to second order in the curvature. 
In effect, up to this order we have in Fourier space (see App.~\ref{app:geometric}) 
\begin{equation}
R^{\mu\nu\rho\sigma}\left(-\mathbf{p}\right)R_{\mu\nu\rho\sigma}\left(\mathbf{p}\right) - 4R^{\mu\nu}\left(-\mathbf{p}\right)R_{\mu\nu}\left(\mathbf{p}\right) + R\left(-\mathbf{p}\right)R\left(\mathbf{p}\right)=0\, ;
\end{equation}
in its turn, integrating over the momenta and using  Parseval's identity, we are lead to the an expression in configuration space,
\begin{equation}
\int {\mathrm{d}^D\mathbf{x}} \left (R^{\mu\nu\rho\sigma}R_{\mu\nu\rho\sigma} - 4R^{\mu\nu}R_{\mu\nu} + R^2 \right)=0\,,
	\label{ec:Gauss_Bonnet}
\end{equation}
which is the Gauss--Bonnet identity for weak gravitational fields in $D$ dimensions.
On the other hand, from the definition of the Weyl tensor we get
\begin{equation}
C^{\mu\nu\rho\sigma}\left(-\mathbf{p}\right)C_{\mu\nu\rho\sigma}\left(\mathbf{p}\right) = R^{\mu\nu\rho\sigma}\left(-\mathbf{p}\right)R_{\mu\nu\rho\sigma}\left(\mathbf{p}\right)- \frac{4}{D-2} R^{\mu\nu}\left(-\mathbf{p}\right)R_{\mu\nu}\left(\mathbf{p}\right)+ \frac{2}{(D-1)(D-2)} R\left(-\mathbf{p}\right)R\left(\mathbf{p}\right).
\end{equation}
Using these results we obtain
\begin{multline}	P =  \frac{\pi^{3/2-D/2}}{4^{D}\Gamma\left(\frac{D+3}{2}\right)} \int\frac{\mathrm{d}^D\mathbf{p}}{(2\pi)^D}\,\Theta\left(-\mathbf{p}^2\right)\left(-\mathbf{p}^2\right)^{\frac{D}{2}-2} \\
	\times \left[ \left(D^2-1\right)\left(\xi - \xi_D\right)^2  R\left(-\mathbf{p}\right)R\left(\mathbf{p}\right) + \frac{D-2}{8\left(D-3\right)} C^{\mu\nu\rho\sigma}\left(-\mathbf{p}\right)C_{\mu\nu\rho\sigma}\left(\mathbf{p}\right) \right].
	\label{ec:proba_final_m00}
\end{multline}
This equation generalizes to arbitrary dimensions the well-known result   obtained in Ref.~\cite{Frieman,Maroto,Campos94} for $D=4$ (see also~\cite{Schubert}). It also shows that, up to second order in the curvature and for $D\geq 4$, there is no particle creation in conformally flat metrics for conformally coupled fields.
If $D\to3$ one obtains an undetermined expression, given that the apparent pole might be compensated by the  trivially vanishing of the Weyl tensor; we will see in Sec.~\ref{sec:cotton} that this is indeed the case.


\subsection{Electric and magnetic part of the Weyl tensor}

We will now write the imaginary part of the effective action in terms of the electric and magnetic parts of the Weyl tensor, showing a complete analogy with the particle creation rate due to electromagnetic fields in the weak field approximation.  
Taking into account the symmetries of the Weyl tensor we have the identity
\begin{equation}	C^{\mu\nu\rho\sigma}C_{\mu\nu\rho\sigma} = 4C^{0 i 0 j}C_{0 i 0 j} + 4C^{0 i j k}C_{0 i j k} + C^{i j k l}C_{i j k l},
	\label{electro:weyl^2_1}
\end{equation}
which is valid in $D$ dimensions.

From now on we consider the physically more relevant case of $D=4$. 
Then the electric and magnetic parts of the Weyl tensor can be defined 
 as \cite{Matte}
\begin{align}
\left\{
\begin{aligned}
	E_{ij} &= \frac{1}{4} \epsilon_{abi}\epsilon_{cdj} C^{abcd},\\
	B_{ij} &= -\frac{1}{2}\epsilon_{i a b} {C_{0 j}}^{a b} ,
\end{aligned}
\right.
\label{electro:def}
\end{align}
where $\epsilon_{ijk}$ is the three-dimensional Levi--Civita pseudotensor (we recall that we are considering weak gravitational fields). These definitions imply that $C_{0i0j} = - E_{ij}$ and 
	$B^{i j}B_{i j} = -\frac{1}{2} C^{0 j a b}C_{0 j a b}$,
which leads to
\begin{equation}
	C^{\mu\nu\rho\sigma}C_{\mu\nu\rho\sigma} = 8 \left( E^{ij}E_{ij} - B^{ij}B_{ij} \right),
\end{equation}
as well as the alternative formula in momentum space
\begin{equation}
 C^{\mu\nu\rho\sigma}\left(-\mathbf{p}\right)C_{\mu\nu\rho\sigma}\left(\mathbf{p}\right) = 8 \left( E^{ij}\left(-\mathbf{p}\right)E_{ij}\left(\mathbf{p}\right)-B^{ij}\left(-\mathbf{p}\right)B_{ij}\left(\mathbf{p}\right)\right).
	\label{electro:weyl^2_final}
\end{equation}

Using this decomposition, the  $D=4$ pair production probability can be rewritten as 
\begin{equation}
	P =  \frac{1}{240\pi} \int\frac{\mathrm{d}^4\mathbf{p}}{(2\pi)^4}\,\Theta\left(-\mathbf{p}^2\right)\left[ \frac{15}{2}\left(\xi - \frac{1}{6}\right)^2R\left(-\mathbf{p}\right)R\left(\mathbf{p}\right) + 
 E^{ij}\left(-\mathbf{p}\right)E_{ij}\left(\mathbf{p}\right)-B^{ij}\left(-\mathbf{p}\right)B_{ij}\left(\mathbf{p}\right) 
 \right].
	\label{electro:proba_final}
\end{equation}
This expression is the gravitational analog of the usual textbook formula  for the imaginary part of effective action
in QED \cite{Itzykson}, which depends on the electric ($\vec{E}$) and magnetic ($\vec{B}$) field  through the invariant
$|\vec{E}\left(-\mathbf{p}\right)\vert^2 - |\vec{B}\left(-\mathbf{p}\right)\vert ^2$. From Eq. \eqref{electro:proba_final} we can conclude that the physical origin of particle creation in conformal field theories lies in the electric part of the Weyl tensor.

\subsection{Particle creation due to an oscillating Newtonian star}

As an example of the previous developments, we will compute the particle creation rate produced by a  star with a time dependent radius, considering  a massless quantum field in $D=4$. As we are working within the weak gravitational field approximation, we will consider a Newtonian star, and the time dependence will be assumed to be nonrelativistic.

We model the spherically symmetric spacetime as
\begin{equation}
{\rm d}s^2 = - \big(1+2\varphi(t,r)\big){\rm d}t^2 +\big(1-2\varphi(t,r)\big)({\rm d}\vec x)^2 \, ,
\label{metrica_estrella}
\end{equation}
 where $r=\vert\vec x\vert$.  Outside a static star, the  potential is Newtonian, i.e. $\varphi(r) = -MG/r$. For a time dependent radius $a(t)$, we will assume that  $\varphi(t,r)$  depends on $t$ only locally through the function $a(t)$, and that it interpolates between a constant at $r=0$ and $-MG/a(t)$ at $r=a(t)$, that is 
 \begin{align}
	\varphi\left(t,r\right)= \begin{cases}
		-\frac{GM}{r}f(r/a(t)) & \text{if } r < a(t) \\
		-\frac{GM}{r} & \text{if } r \geq a(t)\, 
	\end{cases}
	\label{ejemplo2:potencialf}
\end{align}
and the function $f$ depends on the structure of the star.
 In order to avoid singular terms in the curvature, we will assume that the potential and its first and second $r$-derivatives  are continuous at $r=a(t)$, that is  $f(1)=1$, while $f'(1)=f''(1)=0$, where the tilde denotes derivative with respect to the argument.
 
For example,  if we further choose the function $f$
as an odd polynomial of fifth order,
$f(x) = c_1 x +c_2 x^3 +c_3 x^5$, 
the continuity conditions at $r=a(t)$ determine
\begin{align}
	\varphi = \begin{cases}
		-\frac{15GM}{8a}\left(1-\frac{2r^2}{3a^2}+\frac{r^4}{5a^4}\right) & \text{if } r < a \\
		-\frac{GM}{r} & \text{if } r \geq a
	\end{cases}.
	\label{ejemplo2:potencial0}
\end{align}
This is also the Newtonian potential for a spherically symmetric mass $M$ of an ideal gas with an equation of state given by
$P(t,r)=R_g\rho(t,r)T(t,r)$, 
where $R_g$ is the gas constant, while  $P=P(t,r)$,  $\rho=\rho(t,r)$ and $T(t,r)$ are the pressure, density, and temperature, respectively, if one imposes the additional condition at the center of the star $P_0=\kappa \rho_0^{\gamma}$
 ($\kappa>0$ and $\gamma$ are constants that depend on the composition of the star \cite{Jun2000}). When $\gamma > 4/3$,  there exists a unique stationary solution $a_0 = a_0(M, \gamma, \kappa)$ for the radius of the star and for the frequency $\omega = \omega(M, \gamma, \kappa)$ in  the small oscillations regime.

 These details will not be relevant in what follows; indeed, we will work with an arbitrary function $f$ in our calculations, in order to keep track of the eventual dependence of the results with the inner structure of the star.
In particular, assuming non-damped radial oscillations given by $a(t) = a_0 \left(1 + \epsilon\, \text{cos}(\omega t)\right)$, $\epsilon \ll 1$, the Newtonian potential takes the form
\begin{equation}
	\varphi\left(t,r\right) = \varphi_0\left(r\right) + \epsilon \frac{GM}{a_0}f'(r/a_0)\, \text{cos}(\omega t) \,\Theta\left(a_0-r\right) + \mathcal{O}\left(\epsilon^2\right),
	\label{ejemplo2:pot_epsilon}
\end{equation}
where the time-independent function $\varphi_0$ denotes the potential $\varphi$ of Eq. \eqref{ejemplo2:potencialf} with $\epsilon=0$.

 In order to compute the particle creation rate using Eq. \eqref{ec:proba_final_m0} in $D=4$, we will need the explicit expression of the Ricci tensor. 
 For the spacetime metric \eqref{metrica_estrella} and up to linear order in $\varphi$, it reads
 \begin{eqnarray}
R_{00}&=& \square\varphi +4\, \partial_0^2\varphi,\nonumber\\
R_{0i}&=& 2\,\partial_0\partial_i\varphi,\nonumber\\
R_{ij}&=& \square\varphi\, \delta_{ij}\, .
 \end{eqnarray}
Its expression in momentum space can be obtained from
the Fourier transform of Eq. \eqref{ejemplo2:pot_epsilon}, which is given by
\begin{equation}
	\varphi\left(\mathbf{p}\right) = 2\pi\varphi_0(p)\delta\left(p_0\right)\\
+ 4\pi^2 \epsilon \frac{GM}{a_0}\, \big[\delta\left(p^0-\omega\right)+\delta\left(p^0+\omega\right)\big] \int_0^{a_0} {\rm d}r\,  r^2f'(r/a_0)\frac{\sin pr}{pr},
	\label{ec:ejemplo_transformada0}
\end{equation}
being $p=\vert\vec p\vert$ and $\varphi_0(p)$ the spatial Fourier transform of $\varphi_0\left(r\right)$. The first term, associated with the time-independent part of the Newtonian potential, does not contribute to particle creation due to the factor $\Theta\left(-\mathbf{p}^2\right)\delta\left(p_0\right) $ in the integral of  Eq.~\eqref{ec:proba_final_m0}. To proceed further we note that, due to the threshold, $p<\omega$, and therefore $p r\ll 1$ in the nonrelativistic limit $\omega a_0\ll 1$. Thus, in this approximation we obtain
\begin{equation}
	\varphi\left(\mathbf{p}\right) \approx 2\pi\varphi_0(p)\delta\left(p_0\right) + 4\pi^2\epsilon GM a_0^2\, \alpha_f\left[\delta\left(p^0-\omega\right)+\delta\left(p^0+\omega\right)\right],
	\label{ec:ejemplo_transformada}
\end{equation}
where we defined the (structure-dependent) constant
\begin{equation}
    \alpha_f=\int_0^1 {\rm d}x\, x^2f'(x).
\end{equation}
For the particular potential of Eq. \eqref{ejemplo2:potencial0} we have $\alpha_f=1/7$; Other choices of the function $f$  will in general produce different values.
Lastly, inserting Eq. \eqref{ec:ejemplo_transformada} in the expression \eqref{ec:proba_final_m0}, after straightforward calculations we  get the particle production rate
\begin{equation}
	\frac{P}{\tau} = \frac{34}{35\pi} \left(\left(\xi-\frac{1}{6}\right)^2+\frac{1}{612}\right)\left(\epsilon GM \alpha_f\right)^2 a_0^4\,\omega^7
	+O(\omega^9)\, ,
\end{equation}
where $\tau = 2\pi\delta(0)$ represents the total duration of the undamped oscillations.

This example highlights several key aspects of particle creation. First, time-independent terms do not contribute to the imaginary part of the effective action. Second, the particle creation rate reaches a minimum when the field is conformally coupled to the curvature, i.e., \(\xi = \xi_4 = 1/6\); this will become clear for general backgrounds in Sec.~\ref{sec:threshold}, considering Eq.~\eqref{electro:proba_final}. Finally, the rate also depends on the star’s internal structure through the constant \(\alpha_f\), which encodes information about the matter distribution.
The influence of the internal structure on vacuum polarization in (static) Newtonian stars was first established in Ref. \cite{stars1} (see also \cite{stars2}). Vacuum polarization effects during gravitational collapse have also been studied in Ref. \cite{collapse_Calmet} using the real part of the nonlocal effective action. Our results demonstrate that a similar dependence on internal structure extends to the particle creation rate. A deeper analysis beyond the weak gravitational field approximation will be left to future  work.

\section{Particle creation rate and  Cotton tensor}\label{sec:cotton}

As we have previously stated, the final result in Eq.~\eqref{ec:proba_final_m00} is not valid for $D=3$, where the last term in Eq.~\eqref{ec:proba_final_m00} is 
ill defined due to the $1/(D-3)$ factor and the Weyl tensor vanishes identically. 
This is a special case, inasmuch as for $D=3$ a metric is conformally flat iff the Cotton tensor vanishes, whereas for $D\geq 4$ the condition for conformal flatness is the vanishing of the Weyl tensor. This fact suggests
that the pair production probability should be written in terms of the Cotton tensor in $D=3$. 
We will see that this rewriting will apply to any number of dimensions $D\geq 3$.  Notice that this is not in contradiction with our previous claims about the absence of particle creation for conformally flat metrics, since the Cotton tensor vanishes for conformally flat metrics in any number of dimensions (the converse is true only for $D=3$).

Recall then that, in $D$ dimensions, the definition of the Cotton tensor is
\begin{equation}
	C_{\mu\nu\rho} = \nabla_{\rho}R_{\mu\nu} - \nabla_{\nu}R_{\mu\rho} + \frac{1}{2(D-1)}\left( g_{\mu\rho}\nabla_{\nu}R - g_{\mu\nu}\nabla_{\rho}R \right) .
	\label{ec:Cotton_definicion}
\end{equation}
Up to second order in the curvature, this tensor satisfies a Gauss--Bonnet like identity, which in Fourier space reads (see App.~\ref{app:geometric})
\begin{equation}
	 \frac{1}{2\mathbf{p}^2}
  C^{\mu\nu\rho}\left(-\mathbf{p}\right)C_{\mu\nu\rho}\left(\mathbf{p}\right)  
  = R^{\mu\nu}\left(-\mathbf{p}\right)R_{\mu\nu}\left(\mathbf{p}\right)  - \frac{D}{4(D-1)}R\left(-\mathbf{p}\right)R\left(\mathbf{p}\right);
	\label{ec:cotton_identidad}
\end{equation}
this implies that, in configuration space, the following integrated bilinear expression in the curvatures vanishes:
\begin{equation}
	\int \mathrm{d}^D\mathbf{x} 
    \left(
    C^{\mu\nu\rho}C_{\mu\nu\rho} + 2 R^{\mu\nu}\square R_{\mu\nu} - \frac{D}{2(D-1)} R\square R 
    \right)=0\, .
\end{equation}
Inserting the identity~(\ref{ec:cotton_identidad}) into Eq. \eqref{ec:Avramidi_euclidea}
we can write the full effective action  in terms of the Ricci scalar and the Cotton tensor. The result reads
\begin{align}
\begin{split}
\Gamma^{(2)}_{\text{E}} =\frac{1}{2(4\pi)^{D/2}} \int \mathrm{d}^D\mathbf{x}\,\sqrt{g} &\left[
	-\frac{1}{2}C_{\mu\nu\rho} 
	 \frac{1}{\square_E}
	 \beta^{(3)}(\square_E)C^{\mu\nu\rho} 
	\right.\\
	 &\hspace{1cm} \left. +R\left(\xi^2\beta^{(1)}(\square_E)-2\xi\beta^{(2)}(\square_E)+\beta^{(4)}(\square_E) +\frac{D}{4(D-1)}\beta^{(3)}(\square_E)\right)R\right],	\label{ec:Avramidi_euclidea_Cotton}
\end{split}
\end{align}
where we have set $\sigma^2=0$. In particular, for massless quantum fields one can readily show that the coefficient of the nonlocal term that contains the Ricci scalar is proportional
to $(\xi-\xi_D)^2$ in any number of dimensions. Consequently, for conformal fields, the effective action depends exclusively on the Cotton tensor.

Using these results and Wick-rotating to a Lorentzian signature, we obtain from the imaginary part of the effective action  the following probability of pair creation:
\begin{multline}
	P =  \frac{\pi^{(3-D)/2}}{4^{D}\Gamma\left(\frac{D+3}{2}\right)} \int\frac{\mathrm{d}^D\mathbf{p}}{(2\pi)^D}\,\Theta\left(-\mathbf{p}^2\right)\left(-\mathbf{p}^2\right)^{\frac{D}{2}-2}
	\left[ \left(D^2-1\right)\left(\xi - \xi_D\right)^2 R\left(-\mathbf{p}\right)R\left(\mathbf{p}\right) + \frac{1}{4\mathbf{p}^2} C^{\mu\nu\rho}\left(-\mathbf{p}\right)C_{\mu\nu\rho}\left(\mathbf{p}\right) \right].
	\label{ec:proba_final_m0_cotton}
\end{multline}
This expression shows that, in any number of dimensions $D\geq 3$, and up to second order in the curvatures,  there is no particle creation 
for conformally coupled fields in conformally flat metrics. 
The equivalence of Eqs. \eqref{ec:proba_final_m0_cotton}
and \eqref{ec:proba_final_m00} for $D>3$ can alternatively be proved using the identity
\begin{equation}
	\mathbf{p}^2\,  C^{\mu\nu\rho\sigma}\left(-\mathbf{p}\right)C_{\mu\nu\rho\sigma}\left(\mathbf{p}\right)   = 2 \frac{D-3}{D-2}\, C^{\mu\nu\rho}\left(-\mathbf{p}\right)C_{\mu\nu\rho}\left(\mathbf{p}\right)  ,
\end{equation}
a proof of which can be found in App.~\ref{app:geometric}.


\section{Threshold for particle creation and positivity of $\mathrm{Im}(\Gamma)$}\label{sec:threshold}

The particle creation rate for massive fields has the usual threshold
for the pair creation,
which in our convention for the signature of the metric is equivalent to
\begin{equation}
\Theta\left(-4 m^2-\mathbf{p}^2\right)=\Theta\left(p_0^2-\vec{p}\,^2-4m^2\right)\, .
\end{equation}
In particular,  even for massless fields  this implies that the only modes that contribute are those such that $\vert p_0 \vert >\vert\vec p\vert$, i.e., the background fields should depend on time in order to be able to produce particles.

This condition is crucial to ensure that the imaginary part of the effective action is positive definite or, in other words, that the vacuum persistence probability is less than one. This fact is well known and has been used to show that, in QED, particle creation in homogeneous backgrounds is a purely electric effect \cite{Itzykson}.  We will show here that the same holds true in the gravitational counterpart.

The proof goes as follows.
In the weak field approximation  of Eq.~\eqref{eq:expansion}
we have
\begin{equation}
	B_{ij}\left(\mathbf{x}\right) = \frac{1}{2} \epsilon_{iab} \Big[ \partial_{j}\partial_{b}h_{0a}\left(\mathbf{x}\right) + \partial_{a}\partial_{0}h_{jb}\left(\mathbf{x}\right) 
    \Big]
    +
    \frac{1}{4}\epsilon_{i a j }\Big[
    \Box h_{0 a }\left(\mathbf{x}\right) -  \partial_{0 }\partial^{\alpha }h_{a \alpha }\left(\mathbf{x}\right) -  \partial_{a }\partial^{\alpha }h_{0\alpha }\left(\mathbf{x}\right) + \partial_{a }\partial_{0 }h\left(\mathbf{x}\right)
    \Big],
\end{equation}
or, in momentum space,
\begin{equation}
	B_{ij}\left(\mathbf{p}\right) = 
    -\frac{1}{2} \epsilon_{iab} \Big[ p_{j}p_{b}h_{0a}\left(\mathbf{p}\right) + p_{a}p_{0}h_{jb}\left(\mathbf{p}\right) 
    \Big]
    -
    \frac{1}{4}\epsilon_{i a j }\Big[
    \vec p\,^2 h_{0 a }\left(\mathbf{p}\right) -  p_{0 }p_b h_{ab}\left(\mathbf{p}\right) -  p_{a }p^{\alpha }h_{0\alpha }\left(\mathbf{p}\right) + p_{a }p_{0 }h\left(\mathbf{p}\right)
    \Big].
	\label{electro:B_lineal}
\end{equation}
Given that the condition  $\mathbf{p}^2<0$ is tantamount to saying that  $\mathbf{p}$ 
is a timelike four-vector, we can therefore choose a reference system in which $\vec p=0$ and, consequently, $B_{ij}=0$. Moreover, since the effective action is Lorentz invariant,  this implies that $\vert E_{ij}\vert^2-\vert B_{ij}\vert^2 \geq 0$ in any reference frame, and then the pair production probability in  Eq.~\eqref{electro:proba_final} is always positive. Moreover, for a fixed metric,  the probability is minimized by the conformal coupling. This proof can be straightforwardly extended to any number of dimensions, without making reference to the electric and magnetic parts of the Weyl tensor. One can also explicitly check  that, when writing the particle creation rate in terms of the Cotton tensor,
positivity of the imaginary part of the effective action follows from the property (see App.~\ref{app:geometric})
\begin{equation}
\frac{1}{4\mathbf{p}^2} \Theta\left(-\mathbf{p}^2\right)C^{\mu\nu\rho}\left(-\mathbf{p}\right)C_{\mu\nu\rho}\left(\mathbf{p}\right)\geq0\, .
\label{ec:cotton+}
\end{equation}

The existence of the threshold for massless fields, i.e. $\mathbf{p}^2<0$, seems not to have been properly taken into account in some recent works. 
In particular, the results of Ref.~\cite{Wondrak}, where the authors considered massless quantum fields in the presence of classical backgrounds, has sparkled several discussions~\cite{Ferreiro,reply_Wondrak,Schubert,Loeb} 
In light of our findings, we would like to add some additional remarks. Although Ref.~\cite{Wondrak} introduces the full effective action up to second order in the curvatures, at some point the authors ignore the derivatives of the geometric invariants. In this approximation, 
the threshold only depends on the mass;
taking afterwards the massless limit, one is lead to the undetermined value $\Theta(0)$, which is ultimately replaced by one half. 
Taking this replacement into account, the momentum integral that defines the imaginary part of the effective action can be transformed into a spacetime integral of the (squared) curvatures using Parseval's theorem, from which the authors concluded that  static gravitational or electromagnetic  backgrounds may lead to pair creation effects, even in the weak field approximation.  
Note that, according to this line of reasoning, the imaginary part of the effective action would be nonvanishing even for a field with a constant squared mass $\sigma^2=\sigma_0^2$, what is clearly incorrect. 

To understand the source of this fallacy, notice the following.
First of all, we would like to stress that the calculations of the effective action in the in-out formalism assume that, asymptotically, the in and out vacua are well defined, i.e., that the external fields are switched off both far in the future and far in the past (for a recent discussion on this subject see \cite{Akhmedov2024}). 
In addition, the formalism of Barvinsky and Vilkovisky explained in Sec.~\ref{sec:NL} requires an asymptotically flat geometry,
so that the use of these results for arbitrary configurations requires prudence.

Second, we have proved that, analogously to the electromagnetic case~\cite{Ferreiro},  in the presence of gravitational backgrounds the momentum threshold is crucial for unitarity, cf. Eq.~\eqref{electro:proba_final}. In effect, if due attention is not paid to this issue, the imaginary part of the effective action could develop a negative contribution, what would lead to spurious violations of unitarity for classical  gravitational fields dominated by the magnetic component of the Weyl tensor (the vacuum persistence probability could become larger than one).
   
As far as we can see, the origin of the inconsistencies in Ref.~\cite{Wondrak} lies in the fact that the second order effective action derived by Barvinsky and Vilkovisky assumes $\Box\mathcal{R}\gg \mathcal{R}^2$, so that the only remnant scale is symbolically given by $m^2/\Box$ and a naive succession of the limits $\Box\to 0$ and $m\to0$ is simply not allowed.
Indeed, for massless fields, the momentum integral in our expression~\eqref{ec:proba_final_m00} for the pair creation probability contains the factor $\Theta\left(-\mathbf{p}^2\right)$;
as correctly stressed already in Ref.~\cite{Frieman}, this integral can be transformed into a spacetime integral via Parseval's theorem only when the integrand $X$ has a support contained on a subspace of timelike points,
\begin{equation}\label{condit}
     \Theta(-\mathbf p^2) X(\mathbf{p}) =X(\mathbf{p})\, ,
\end{equation} 
which is not satisfied for static fields.
This objection seems also to have been overlooked in other works \cite{Chernodub,Banerjee}. For instance, in Ref.\cite{Chernodub} it was pointed out that the imaginary part for the effective action  is essentially the spacetime integral of the Schwinger--DeWitt $a_2$ heat-kernel coefficient and, therefore, would be related  to the trace anomaly. We insist then that these results could be valid only when the condition in Eq.~\eqref{condit} is satisfied.

The confusion that static backgrounds can induce the creation of pairs may be attributed to the fact that, when considering the paradigmatic example of the Schwinger effect, one customarily refers to a static electric field. However, in Schwinger's calculation it is also implicitly assumed that the electric field is switched off for $t\to\pm\infty$ (for an enlightening discussion that includes particle creation in  de Sitter space see Ref.~\cite{AndersonMottola}). 
Leaving this aside for the moment, there are several ways to understand Schwinger's results. 

On the one hand, in a time-dependent gauge, the electric field is derived from a time-dependent potential vector. The modes of a charged field do depend on the potential vector, and one can compute a nontrivial Bogoliubov transformation
between the in and out vacua, which leads to the creation of pairs. 

On the other hand, when considering a static gauge, though the notion of positive frequency modes  $\exp[-i\omega t]$ is well defined,  in reality the theory has no ground state. This has already been emphasized in Ref.~\cite{Padmanabham91}, where it is also pointed out that,  even for a single harmonic oscillator, the imaginary part of the effective action could differ from zero 
both for
 a non adiabatic time evolution of the frequency and for an adiabatic evolution in which the squared frequency becomes negative. The latter is the case for a static electric field in a static gauge,
 where the associated equations for the modes of the quantum field display the analogue of a negative squared mass $\sigma^2(x)=-\sigma_0^2 x^2$, $\sigma_0\in\mathbb{R}$. In other words, although in a static situation one can always choose positive frequency solutions, the associated ``vacuum" state does not minimize the Hamiltonian and is unstable.
This instability distinguishes the constant electric field background from other classical time-independent backgrounds.

In the gravitational case, such instability is not present for weak fields. Indeed, explicit calculations of $\langle T_{\mu\nu}\rangle$ around Newtonian stars give a $1/r^5$ decay at infinity, i.e., there is no energy  flux that one could associate to 
particle creation~\cite{stars1,stars2}. Therefore, a static, weak gravitational field does not produce particles. 
The presence of event horizons
may invalidate this conclusion, since the vacuum state associated to the Killing vector may have an unphysical divergence on the horizon, as is the case for the Boulware vacuum in the Schwarzschild metric
\cite{ChristFull}. 

\section{Discussion}\label{sec:discussion}

In this paper we have analyzed the particle creation effect induced by general weak gravitational fields, in an arbitrary number of dimensions. We have derived the vacuum persistence probability from the imaginary part of the effective action, which have been calculated up to second order in the curvature using the so-called covariant perturbation theory.

Despite the structural differences between the nonlocal operators present in the effective action for even and odd dimensions, the imaginary part admits a unified description valid for arbitrary dimensions. This analysis is summarized in Eq.~\eqref{ec:proba_final}, which involves the Fourier transforms of the background fields and the essential threshold for particle creation.

For massless fields in \( D \geq 4 \), the master formula for the imaginary part of the effective action is Eq.~\eqref
{ec:proba_final_m00}. 
It consists of one term proportional to \( (\xi - \xi_D)^2 R^2\) and another proportional to the square of the Weyl tensor. This demonstrates that, up to second order in the curvature, no particle creation occurs for massless, conformally coupled fields when the metric is conformally flat.

Curiously, the general result in Eq.~\eqref
{ec:proba_final_m00} is not well-defined for \( D = 3 \), where it displays a pole. However, we have presented the alternative expression~\eqref{ec:proba_final_m0_cotton}, written in terms of the Cotton tensor, whose range of validity is extended to \( D \geq 3 \). Given that in \( D = 3 \) the Cotton tensor plays the role of the Weyl tensor, i.e. it vanishes for (and only for) conformally flat metrics, our result confirms that there is no particle creation for massless, conformally coupled fields, when the metric is conformally flat in $D=3$.
It is also worth noting that not only the imaginary part of the effective action can be expressed in terms of the Cotton tensor, but the full effective action as well [cf. Eq.~\eqref{ec:Avramidi_euclidea_Cotton}]. This reformulation of the effective action matches the one derived in previous works up to quadratic order in the curvature but deviates beyond that. Consequently, it may be particularly relevant for discussing effective nonlocal modifications to General Relativity from a phenomenological perspective \cite{Maggiore2017}. 

For \( D = 2 \), the situation deserves a separate discussion. In effect, for conformal fields the exact effective action is given by Polyakov's action, which includes an imaginary part at second order in the curvature. This stems from the fact that the trace anomaly is linear in the curvature, so that the second-order effective action already encapsulates anomalous effects. In higher dimensions, we expect that an effective action capturing the conformal anomaly would also lead to particle creation, as is the case for Riegert's action in \( D = 4 \), which is of quartic order in the curvature~\cite{Riegert}.
It would be also compelling to analyze the link between pair creation and the universal quantity derived in Ref.~\cite{Ferrero:2023unz}, given the connection of the latter with the trace anomaly.

Finally, we have emphasized the importance of a threshold being present in the imaginary part of the effective action, as well as its consequences for particle creation. 
Even though in other approaches the threshold is not apparent, such as in Schwinger's exact calculation of the Euler--Heisenberg effective action for static, homogeneous electromagnetic fields~\cite{Heisenberg:1936nmg, Schwinger}, its physical information is equivalently encoded in the analytic structure of the form factors, as a footprint of the primordial instability.
When using the Barvinsky--Vilkovisky resummation, the threshold enforces that weak backgrounds must have nonvanishing Fourier components for timelike momenta, i.e. for $\mathbf{p}^2 
 <0$, if one desires to observe pair production of massless fields. Static backgrounds do not meet this requirement and, therefore, cannot lead to particle creation, unless they render the theory unstable.

\section* {Acknowledgments} We would like to thank Silvia Pla, Jérémie Quevillon, Gonzalo Torroba and Vincenzo Vitagliano for useful discussions. SAF thanks the members of LAPTh, Annecy, for their warm hospitality.
This research was supported by  Consejo Nacional de Investigaciones Científicas y Técnicas (CONICET) and Universidad Nacional de Cuyo (UNCuyo). 
 SAF acknowledges the support from: Helmholtz-Zentrum Dresden-Rossendorf, UNLP through Project 11/X748,  Next Generation EU through the project ``Geometrical and Topological effects on Quantum Matter (GeTOnQuaM)'' and INFN Research Project QGSKY. The Authors extend their appreciation to the Italian National Group of Mathematical Physics (GNFM, INdAM) for its support.

\appendix 

\section{Geometric identities}\label{app:geometric}

We write the metric tensor as
$g_{\mu\nu} = \eta_{\mu\nu} + h_{\mu\nu}$, 
with $\left|h_{\mu\nu}\right|\ll 1$. To first order in the perturbation we have
the expansions
\begin{align}
R_{\alpha\beta\gamma\delta} &= \frac{1}{2}\partial_{\gamma}\partial_{\left[\beta\right.}h_{\left.\alpha\right]\delta} + \frac{1}{2}\partial_{\delta}\partial_{\left[\alpha\right.}h_{\left.\beta\right]\gamma}, \\
R_{\alpha\beta} &= -\frac{1}{2} \partial^{\mu}\partial_{\mu}h_{\alpha\beta} - \frac{1}{2} \partial_{\alpha}\partial_{\beta}h + \frac{1}{2} \partial^{\mu}\partial_{\left(\alpha\right.}h_{\left.\beta\right)\mu},  \\
R &= -\partial^{\mu}\partial_{\mu} h + \partial^{\mu}\partial^{\nu}h_{\mu\nu} ,
\end{align}
where  $h:=\eta^{\mu\nu}h_{\mu\nu}$ and  $[\cdots]$ and $(\cdots)$ respectively denote antisymmetrization and symmetrization in the indices between the brackets. 
Using these expansions, in Fourier space we obtain
\begin{equation}
	R\left(\mathbf{p}\right) = \mathbf{p}^2 h\left(\mathbf{p}\right) - \mathbf{p}^{\mu}\mathbf{p}^{\nu}h_{\mu\nu}\left(\mathbf{p}\right) ,
\end{equation}
and similar expressions for 
$R_{\alpha\beta\gamma\delta}\left(\mathbf{p}\right)$ and $R_{\alpha\beta}\left(\mathbf{p}\right)$.
From these expressions, it is straightforward to check that
\begin{equation}
	R^{\alpha\beta\gamma\delta}\left(-\mathbf{p}\right)R_{\alpha\beta\gamma\delta}\left(\mathbf{p}\right) - 4R^{\alpha\beta}\left(-\mathbf{p}\right)R_{\alpha\beta}\left(\mathbf{p}\right) + R\left(-\mathbf{p}\right)R\left(\mathbf{p}\right) = 0\, .
	\label{ape:gaussbonnet_lin}
\end{equation}
We stress that the result in Eq.~\eqref{ape:gaussbonnet_lin} is valid up to second order in the curvatures in any number of dimensions.
In particular, this is the weak field version of the Gauss-Bonnet theorem, which states that in $D=4$ the integral of 
\begin{equation}
R^{\alpha\beta\gamma\delta}\left(\mathbf{x}\right)R_{\alpha\beta\gamma\delta}\left(\mathbf{x}\right) - 4R^{\alpha\beta}\left(\mathbf{x}\right)R_{\alpha\beta}\left(\mathbf{x}\right) + R^2\left(\mathbf{x}\right)
\end{equation}
over the whole space is a topological invariant.   

We now turn to discuss some useful identities relating the Cotton and Weyl tensors, which are valid for weak gravitational fields.
The Cottton tensor in $D$ dimensions is defined as 
\begin{equation}
	C_{\mu\nu\rho} := \nabla_{\rho}R_{\mu\nu} - \nabla_{\nu}R_{\mu\rho} + \frac{1}{2(D-1)}\left( g_{\mu\rho}\nabla_{\nu}R - g_{\mu\nu}\nabla_{\rho}R \right) \, .
	\label{ape:Cotton_definicion}
\end{equation}
For weak fields in momentum space it thus reads
\begin{equation}
	C_{\mu\nu\rho}\left(\mathbf{p}\right) = i\left( p_{\rho}R_{\mu\nu}\left(\mathbf{p}\right) - p_{\nu}R_{\mu\rho}\left(\mathbf{p}\right) + \frac{1}{2(D-1)}\left( \eta_{\mu\rho}p_{\nu} - \eta_{\mu\nu}p_{\rho}\right) R\left(\mathbf{p}\right) \right),
\label{ec:cottonp}
\end{equation}
where $R_{\mu\nu}\left(\mathbf{p}\right)$ and $R\left(\mathbf{p}\right)$ are evaluated to first order in the perturbation $h_{\mu\nu}$. 
Using the (linearized) contracted Bianchi identity,
$p^{\mu}R_{\mu\nu}\left(\mathbf{p}\right) = \frac{1}{2} p_{\nu}R\left(\mathbf{p}\right)$, an explicit calculation gives a weak field expression of the squared Cotton tensor as a combination of quadratic terms in the Ricci tensor and the Ricci scalar:
\begin{equation}
	C^{\mu\nu\rho}\left(-\mathbf{p}\right)C_{\mu\nu\rho}\left(\mathbf{p}\right) - 2\mathbf{p}^2 R^{\mu\nu}\left(-\mathbf{p}\right) R_{\mu\nu}\left(\mathbf{p}\right) + \frac{D}{2(D-1)} \mathbf{p}^2 R\left(-\mathbf{p}\right) R\left(\mathbf{p}\right) = 0.
 \label{eq:Cottonid}
\end{equation}

On the other hand, the squared Weyl tensor in momentum space is given by
\begin{align}
	C_{\mu\nu\rho\sigma} \left(-\mathbf{p}\right)C^{\mu\nu\rho\sigma} \left(\mathbf{p}\right)&= R_{\mu\nu\rho\sigma} \left(-\mathbf{p}\right)R^{\mu\nu\rho\sigma} \left(\mathbf{p}\right) - \frac{4}{D-2} R_{\mu\nu} \left(-\mathbf{p}\right)R^{\mu\nu} \left(\mathbf{p}\right) + \frac{2}{(D-1)(D-2)} R\left(-\mathbf{p}\right)R\left(\mathbf{p}\right)
    \notag \\
	&= \frac{4(D-3)}{D-2}R_{\mu\nu} \left(-\mathbf{p}\right)R^{\mu\nu} \left(\mathbf{p}\right)  - \frac{D(D-3)}{(D-1)(D-2)} R\left(-\mathbf{p}\right)R\left(\mathbf{p}\right) ,
	\label{ape:relacion0}
\end{align}
where in the second equality we used Eq.~\eqref{ape:gaussbonnet_lin}.
Inserting the identity~\eqref{eq:Cottonid} into Eq.~\eqref{ape:relacion0},
we arrive at 
\begin{equation}
	\mathbf{p}^2 C_{\mu\nu\rho\sigma} \left(-\mathbf{p}\right)C^{\mu\nu\rho\sigma} \left(\mathbf{p}\right) = 2 \frac{D-3}{D-2} C^{\mu\nu\rho}\left(-\mathbf{p}\right)C_{\mu\nu\rho}\left(\mathbf{p}\right),
	\label{ape:relacion00}
\end{equation}
which is valid for $D\geq 3$ (recall that the Weyl tensor trivially vanishes for $D=3$).

Finally, we provide a proof for Eq.~\eqref{ec:cotton+}.
 To this end, we expand the sum defining the squared Cotton tensor in momentum space, distinguishing between terms with an even and an odd number of temporal indices (remember that, in the weak field approximation, the indices are lowered with $\eta_{\mu\nu}$):
\begin{multline}
    C^{\mu\nu\rho}\left(-\mathbf{p}\right)C_{\mu\nu\rho}\left(\mathbf{p}\right) = \underbrace{2C^{00i}\left(-\mathbf{p}\right)C_{00i}\left(\mathbf{p}\right) + C^{ijk}\left(-\mathbf{p}\right)C_{ijk}\left(\mathbf{p}\right)}_{\geq 0} \\
    \underbrace{+ C^{0ij}\left(-\mathbf{p}\right)C_{0ij}\left(\mathbf{p}\right) + 2C^{ij0}\left(-\mathbf{p}\right)C_{ij0}\left(\mathbf{p}\right) }_{\leq 0}.
\label{ec:cotton_descomposicion}
\end{multline}
We will see that the positive terms in (\ref{ec:cotton_descomposicion}) are always proportional to some spatial component of $\mathbf{p}$. 
From Eq.~\eqref{ec:cottonp}
we have
\begin{equation}
C_{ijk}\left(\mathbf{p}\right) = i\left[ p_{k}R_{ij}\left(\mathbf{p}\right) -  p_{j}R_{ik}\left(\mathbf{p}\right) + \frac{1}{2(D-1)}\Big( \eta_{ik} p_{j}R\left(\mathbf{p}\right) - \eta_{ij}  p_{k}R\left(\mathbf{p}\right) \Big) \right],
    \label{ec:Cotton_ijk}
\end{equation}
so this property is evident. On the other hand, 
the components of the Cotton tensor with two temporal indices read
\begin{equation}
    C_{00i}\left(\mathbf{p}\right) = i\left[ p_{i}R_{00}\left(\mathbf{p}\right) -  p_{0}R_{0i}\left(\mathbf{p}\right) + \frac{1}{2(D-1)}  p_{i}R\left(\mathbf{p}\right) \right].
    \label{ec:Cotton_00i}
\end{equation}
Using the expression for $R_{0i}\left(\mathbf{p}\right)$, one can readily check that, once again, all terms are proportional to  some
spatial component of $\mathbf{p}$.
Due to the threshold, 
$\mathbf{p}$ is a time-like vector; thus, we can choose an inertial frame in which the spatial components of $\mathbf{p}$ vanish. So, in this frame we have
 $C^{\mu\nu\rho}\left(-\mathbf{p}\right)C_{\mu\nu\rho}\left(\mathbf{p}\right)\leq 0$. As this is a Lorentz invariant quantity, the inequality holds for any inertial frame. Therefore, we have proved Eq.~\eqref{ec:cotton+}:
\begin{equation}
    \underbrace{\frac{1}{4\mathbf{p}^2}}_{\leq 0} \Theta\left(-\mathbf{p}^2\right) \underbrace{C^{\mu\nu\rho}\left(-\mathbf{p}\right)C_{\mu\nu\rho}\left(\mathbf{p}\right)}_{\leq 0} \geq 0.
\end{equation}

\end{document}